\documentclass[journal,10pt]{IEEEtran}

\usepackage{cite}
\usepackage{amsmath,bm}
\usepackage{amssymb}    
\usepackage{amsfonts}
\usepackage{graphicx}    
\usepackage{cases}
\usepackage{color}

\makeatletter
\renewcommand\normalsize{%
   \@setfontsize\normalsize\@xpt\@xipt
   \abovedisplayskip 0.5\p@ \@plus1\p@ \@minus2.5\p@                 
   \abovedisplayshortskip 0\z@ \@plus1\p@ \@minus2.5\p@
   \belowdisplayshortskip 1\p@ \@plus1\p@ \@minus2.5\p@
   \belowdisplayskip \abovedisplayskip
   \let\@listi\@listI}
\makeatother

\ifCLASSINFOpdf

\else

\fi
\begin{document}
\title{ Asymptotically Optimal Estimation Algorithm for the Sparse Signal with Arbitrary Distributions }

\author{\IEEEauthorblockN{Chongwen Huang, \emph{Student Member, IEEE,} Lei Liu, \emph{Student Member, IEEE,} \\ Chau Yuen, \emph{Senior Member, IEEE} }
\thanks{ Copyright (c) 2018 IEEE. Personal use of this material is permitted. However, permission to use this material for any other purposes must be obtained from the IEEE by sending a request to pubs-permissions@ieee.org.

The research of Prof.Yuen was supported by NSFC 61750110529 Grant, and the work of C. Huang was PHC Merlion PhD program.

Chongwen Huang and Chau Yuen are with the Singapore Unversity of Technology and Design, Singapore. Lei Liu is with the Singapore Unversity of Technology and Design and City University of Hong Kong, Hong Kong, China. (e-mail: chongwen$\_$huang@mymail.sutd.edu.sg, leiliuxidian@gmail.com, and yuenchau@sutd.edu.sg).}
}
\maketitle
\begin{abstract}
In this paper, we propose a sparse signal estimation algorithm that is suitable for many wireless communication systems, especially for the future millimeter wave and  underwater communication systems. This algorithm is not only asymptotically optimal, but also robust to the distribution of non-zero entries of the sparse signal. Then, we derive its upper bound and lower bound, and show that the Mean Square Error (MSE) of the proposed algorithm can approach the Minimum Mean Square Error (MMSE) bound when the Signal Noise Ratio (SNR) goes to infinite or zero. Numerical simulations verify our theoretical analysis and also show that the proposed algorithm converges faster than existing algorithms, e.g., TSR-DFT, AMP, etc.
\end{abstract}

\IEEEpeerreviewmaketitle
\section{Introduction}
Sparse signal estimation has gained increasing interest since many wireless communication systems, e.g., millimeter wave ultra-wideband transmission and underwater acoustic communication systems, are best modeled as sparse due to the severe blockage effect and path-loss \cite{Alk_channel_Estimation,SP_for_2016,sparse_channel1,UB_MM}. Specifically, in this paper, we consider the problem of estimating a $L$-sparse signal ($\mathbf{x} \in \mathbb{C}^{N \times 1} $) from the $M < N$ linear measurements that are blended with the noise,
\begin{equation}\label{I1}
\begin{split}
\mathbf{y}= \mathbf{H}\mathbf{x}+ \mathbf{w}, \mathbf{w} \in \mathbb{C}^{M \times 1},
\end{split}
\end{equation}
where $\mathbf{H} \in \mathbb{C}^{M \times N}$  is the known measurement matrix, $\mathbf{y} \in \mathbb{C}^{M \times 1}$ is the observation vector, and $\mathbf{w} \sim \mathcal{N}(0,\sigma^2_{w}\mathbf{I}_{M})$ is the Additive White Gaussian Noise (AWGN).  For the entry $x_i\;( i \in \{1,2,...,N\})$ of the $L$-sparse signal $\mathbf{x} $, it is also denoted as
\begin{equation}\label{I2}
    x_i \sim \left\{\begin{array}{lr}
         0, \quad\;\;\; p(x_i=0)= 1-\lambda \\
       f(x), \;\; p(x_i \neq 0)= \lambda
        \end{array}\right.
\end{equation}
where  $\lambda =L/N$ denotes the sparsity ratio, $p(x_i=0)$ denotes the probability of zero entries in vector $\mathbf{x}$, $p(x_i \neq 0)$ denotes the probability of non-zero entries in vector $\mathbf{x}$, and $f(x)$ is active-coefficient Probability Distribution Function (PDF) that can be any distribution, e.g., Gaussian Mixture distribution, Chi-squared distribution, etc.

For the sparse signal estimation problem, \cite{LASSO} proposed a well-known method named the LASSO. It is possible to obtain the accurate estimation preformation with a low complexity  by using the LASSO, especially when the $\mathbf{x}$ is sufficiently sparse, but this depends on that $\mathbf{H}$  satisfies certain restricted isometry properties \cite{CS,iid_H_Monta}. In additional, the solution of LASSO-type algorithm is generally not the globally optimal.

In some scenarios,  e.g., millimeter wave (mmWave) transmission systems, the sparse signal $\mathbf{x} $  denotes the frequency-domain
impulse responses of the mmWave MIMO channel and can be modeled as independent and identically distributed (i.i.d) Bernoulli-Gaussian (BG) distribution \cite{Alk_channel_Estimation,SP_for_2016,UB_MM}. In other words, the entry $x_i$ is either a zero element, or a non-zero element of the Gaussian distribution. Under such typical case, \cite{majunjie_01} proposed a Turbo-type Signal Recovery algorithm with a partial Discrete Fourier Transform matrices (TSR-DFT). Although it can outperform approximate message passing (AMP)\cite{Donoho_AMP,MP01} that is a low-complexity iterative Bayesian algorithms. The performance of TSR-DFT was only reported under the sparse signal $\mathbf{x} $ that is the i.i.d. BG distribution. Another general scenario is that the sparse signal $\mathbf{x} $ is modeled as the Bernoulli-Gaussian Mixture (GM). For such scenario, there are several algorithms that have been proposed and analysed, e.g., EM-GM-AMP \cite{P_Schniter_EM} algorithm based on AMP and Expectation\text{-}Maximization.

In addition, a ``typical estimator'' that was proposed in \cite{CRB_compressed_sensing1} can asymptotically approach the Genie-aided Minimum Mean Square Error (MMSE) of sparse signal $\mathbf{x}$ under certain constraints on $\mathbf{H}$, but it is from the theoretical perspective. In addition, according to the best of our knowledge, there is no implementable or asymptotically implementable Genie-aided MMSE  estimator that has been presented in the literature, where implementable means that the algorithm can be implemented in practical system.

In this paper, we present a novel turbo-type iteration algorithm that leverages both virtues of the Sparse Message Passing (SMP) \cite{Chongwen_01} and Linear Minimum Mean-Square Error (LMMSE) algorithms for the sparse signal estimation problem. There are main three processes in the proposed algorithm. The first process is to estimate the locations of non-zero elements by SMP, while the second process is to estimate the value of these non-zero elements by leveraging LMMSE and the estimated sparse information in the first process. The first process and second process will help each other at each iteration for improving the performance until the performance of algorithm meets the system requirement. The third process is named as sparsity combiner that is to make decision based on the estimated results in previous two processes. Compared with our previous work \cite{Chongwen_01} and existing work, the main contributions of this paper are summarized as follows.
\begin{itemize}
  \item We design a new method that is robust to the distribution of non-zero entries of the sparse signal $\mathbf{x}$. As compared to the previous work \cite{Chongwen_01} based on the Least Squares Estimator (LSE) that is limited to the deterministic sparse signal $\mathbf{x}$ and lower noise cases, the newly proposed algorithm employs LMMSE that is robust to the distribution of non-zero elements of the signal $\mathbf{x}$.
  \item We derive the lower bound and upper bound of the proposed algorithm, and prove that its performance can approach the MMSE Bound when the Signal Noise Ratio (SNR) goes to infinite or zero. Numerical simulations verify our theoretical analysis and also show that the proposed algorithm converges faster than existing algorithms, e.g., TSR-DFT, AMP, etc.
\end{itemize}

\begin{figure}
  \begin{center}
  \includegraphics[width=84mm]{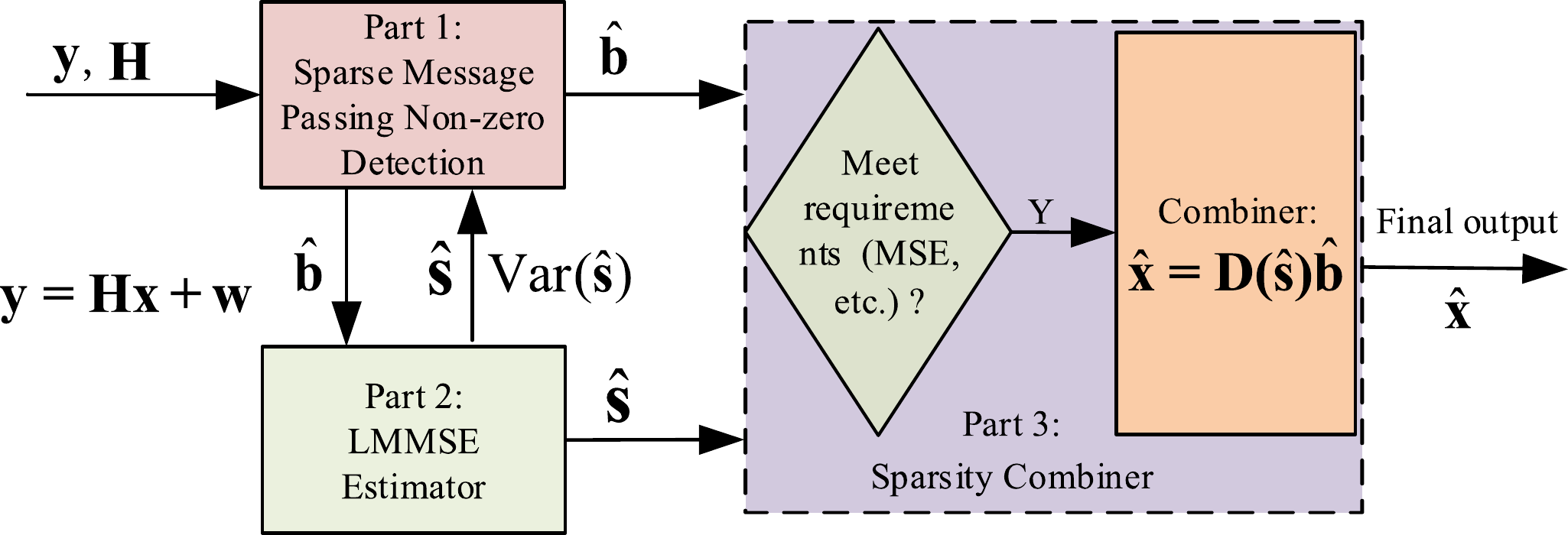}  \vspace{-2mm} %
  \caption{The diagram of the proposed estimation algorithm.  }
  \label{fig:Estimation_Scheme} \vspace{-6mm}
  \end{center}
\end{figure}

\section{SPARSE SIGNAL ESTIMATION}
In this subsection, a novel signal estimation algorithm is introduced as shown in the Fig. \ref{fig:Estimation_Scheme}. This proposed algorithm was named as SMP-LMMSE since it can take full advantage of both virtues of Sparse Message Passing (SMP) and LMMSE. It involves three processes: Sparse Message Passing Non-zero Detection, LMMSE Estimation, and Sparsity Combiner. The Sparse Message Passing Non-zero Detection (Process 1) is to obtain the sparsity information based on the message passing algorithm, sparse feature in the equation (2) and the law of large numbers. Therefore, it also is robust to the distribution of non-zero elements of the signal $\mathbf{x}$. The LMMSE estimator (Process 2) refines the estimate of $\mathbf{x}$ by leveraging the estimated sparsity information in the previous Process 1. Compared with the LSE-SMP estimator in our previous work \cite{Chongwen_01}, where $\mathbf{x}$ is assumed as the deterministic, the Process 2 is also more robust to randomly sparse signal $\mathbf{x}$ and its the distribution of non-zero entries. The first process and second process will iterate each other until the number of iterations reaches the limit or the MSE meets the system requirement, etc. At the end of the iteration, the Sparsity Combiner (Process 3) is to make the decision and output the final estimation $\mathbf{x}$ based on the estimated results of the Process 1 and Process 2. Before we show the detailed operations of the proposed algorithm, we have the following assumption. 

\textbf{\textit{Assumption 1:}} We have the priori information $K$, and the elements of $\mathbf{s}$ and $\mathbf{b}$ are both i.i.d.
\subsection{Sparse Message Passing Non-zero Detection}
Firstly, we focus on the Process 1 and present the novel sparse message passing non-zero detection algorithm to obtain the sparsity information of $\mathbf{x}$, which means to find the positions of non-zero entries of $\mathbf{x}$. The detailed operations of Process 1 will be discussed as follows.
\subsubsection{Factor Graph Representation of the Sparse Signal}
\begin{figure*}
  \begin{center}
  \includegraphics[width=155mm]{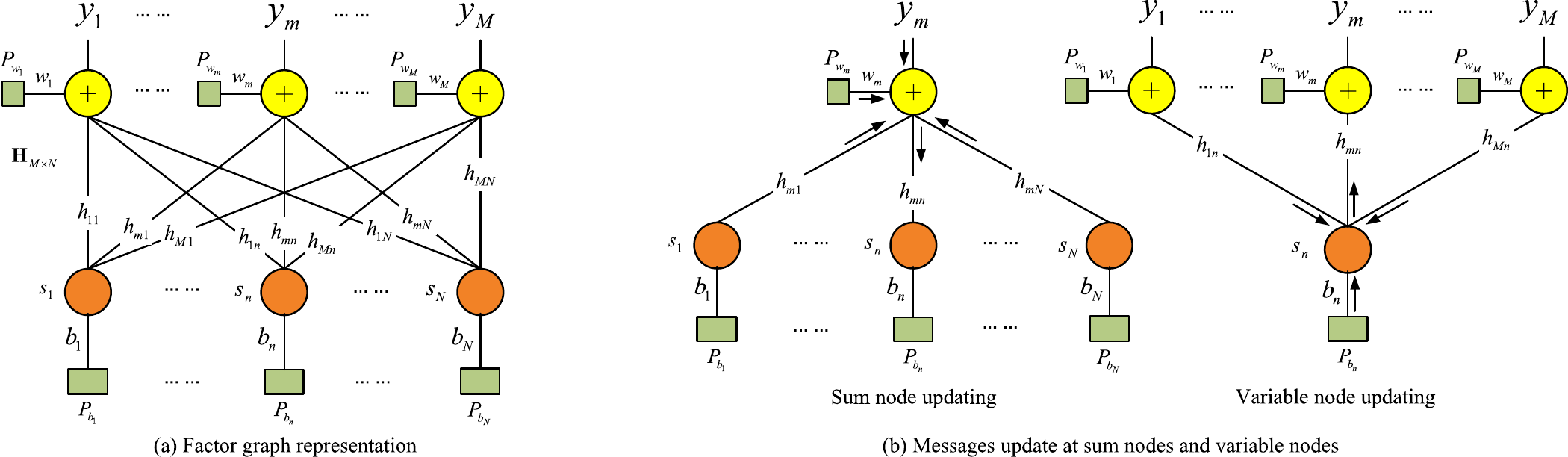} \vspace{-2mm}  \\%
  \caption{ (a) The factor graph representation for the sparse signal $\mathbf{x}$. This representation is based on the expressions (\ref{II2_1})-(\ref{II3_1}) and factor graph rules. (b) Messages update at $m$th sum node and $n$th variable node.  The rule for the message update is that the output message on each edge is updated from the messages on the other edges. Furthermore, the message update on each edge is the probability of the $b_i=1$. The mean and variance of the equivalent Gaussian distribution $n_{mn}^*$ are calculated at the sum nodes according to the equation (\ref{equ_noi}), and they will be used for updating the probability of the $b_i=1$ in next iteration. }
  \label{fig:All_Factor_Graph_L} \vspace{-6mm}
  \end{center}
\end{figure*}

The Fig. \ref{fig:All_Factor_Graph_L} illustrates the sparse message passing non-zero detection algorithm. Specifically, the representation of the sparse signal $\mathbf{x}$  is introduced as the Fig. \ref{fig:All_Factor_Graph_L} (a). Let $\mathbf{A}_{M\times N} \boldsymbol{\cdot} \mathbf{B}_{M\times N}=\left[ a_{ij}b_{ij}\right]_{M\times N}$.  Since $\mathbf{x}$ is the sparse vector as shown in the (2), we define
\begin{equation}\label{II2_1}
\mathbf{x}\triangleq \mathbf{f(\mathbf{s})}\boldsymbol{\cdot} \mathbf{b}=\mathbf{D(\mathbf{s})} \mathbf{b},
\end{equation}
where $\mathbf{s}=[s_1,\cdots,s_{N}]^T$, $\mathbf{b}=[b_1,\cdots,b_{N}]^T$, $\mathbf{D(s)} \in \mathbb{C}^{N \times N}$, and $\mathbf{D(s)}=\mathrm{diag}(\mathbf{s})$ is a diagonal matrix with the elements of $\mathbf{s}$ on its diagonal. Notes that the random vector $\mathbf{b}$ can be denoted as Bernoulli distribution $\mathbf{b}\sim \mathcal{B}^N(1,\lambda)$. Then, when the $b_i=1$, $s_i$ can be seen as the corresponding non-zero elements of $\mathbf{x}$. On the other hand, when the $b_i=0$, $s_i$ can be seen as the corresponding zero elements. Therefore, we rewritten (1) as
\begin{equation}\label{II3_1}
\begin{split}
\mathbf{y}= \mathbf{H}\mathbf{D(\mathbf{s})} \mathbf{b}+ \mathbf{w}.
\end{split}
\end{equation} \vspace{-1mm}
 \vspace{-2mm}

Moreover, we employ the factor graph to represent the equations (\ref{II2_1})-(\ref{II3_1}) based on the typical factor graph update rules \cite{Loeliger_factor_graph,message_passing}, and this result can be seen in the Fig. \ref{fig:All_Factor_Graph_L}. (a). This representation consists of three parts, the sum nodes $ w_1,...,w_M $, variable nodes $ s_{1},...,s_{N} $ and weighted information edges $ h_{11},...,h_{MN} $. The objective of the proposed SMP algorithm is to obtain the exact positions of non-zero elements of $\mathbf{x}$. The rule for the message update is that the output message on each edge is updated from the messages on the other edges\cite{Loeliger_factor_graph,Chongwen_01}. The paper \cite{Lei2015_TWC} has already shown that the massage passing based algorithm can converge to the LMMSE under the Gaussian
distribution. Therefore, this process can obtain near optimal performance for estimating of the sparse information without heavy computational complexity.

\subsubsection{Message Update at Sum Nodes }
The message update in the sum node $m$ can be seen as a multiple-access process as shown in the left part of Fig. \ref{fig:All_Factor_Graph_L} (b) \cite{LDPC_Guassion,message_passing_TVT,Lei2015_TWC}. This shows an example of the sparse message update from the $m$th sum node to the $n$th variable node. Then, the received signal $y_m$ at the $m$th sum node can be denoted as \vspace{1.5mm}
\begin{eqnarray}
y_m = \underbrace{ h_{mn}s_nb_n}_{\mathrm{Desired}} \quad + \underbrace{\sum\limits_{i\in \mathcal{N}/n} {h_{mi}s_{i}b_i} + n_m}_{{\mathrm{Equivalent\, Gaussian \,noise}: \,n^{\ast}_{mn}}},
\end{eqnarray}
where $m\in\mathcal{M}$, $\mathcal{M}=\{1,\cdots, M\}$, and $i\in \mathcal{N}/n$ denotes $n, i \in \mathcal{N} \;\mathrm{and}\; i\neq n$, $\mathcal{N}= \{1,\cdots, N\}$. We assume that $p_{n\to m}^v(\tau)$ denotes the non-zero possibility of the Bernoulli variable $b_n$ passing from the $n$th variable node to the $m$th sum node at the $\tau$th iteration. From the assumption 1, we know that the $s_i, i\in \mathcal{N}/n$ and $ b_i$ are independent with each other. Then, the sum term of  $\sum_{i\in \mathcal{N}/n} {h_{mi}s_{i}b_i} + n_m$ can be approximated as an Equivalent Gaussian Noise $n_{mn}^*$ based on the Law of Large Numbers when the $N$ approximates infinite. Therefore, we have
\begin{equation}
y_m = h_{mn}s_nb_n + n_{mn}^*,\quad n_{mn}^*\sim\mathcal{N}\big(u_{m \to n}^s,v_{m \to n}^s\big),
\end{equation}
where $u_{m \to n}^s$ represents the mean of the $n_{mn}^*$ when the message $ {h_{mn}s_{n}b_n}$ updates from the $m$th sum node to $n$th variable node, and $v_{m \to n}^s$ represents the variance of the $n_{mn}^*$ when the message $ {h_{mn}s_{n}b_n}$ updates from the $m$th sum node to $n$th variable node. In the $\tau$th iteration, the mean $u_{m \to n}^s(\tau)$ and variance $v_{m \to n}^s(\tau)$ of the equivalent Gaussian noise $n_{mn}^*(\tau)$ are derived as follows,
\begin{equation}\label{equ_noi}
\left\{ \begin{array}{ll}
\begin{split}
u_{m \to n}^s(\tau) &= \sum\limits_{i \in {\cal N}/n} {h_{mi}}\hat{s}_i(\tau)p_{i \to m}^v(\tau),\\
v_{m \to n}^s(\tau) &= \sum\limits_{i \in {\cal N}/n} h_{mi}^2p_{i \to m}^v(\tau) { v_{\hat{s}_{i}}(\tau)} + \\
&{h_{mi}^2p_{i \to m}^v(\tau)(1 - p_{i \to m}^v(\tau))\hat{s}_i^{2}(\tau)}  + \sigma _w^2,
\end{split}
\end{array} \right.
\end{equation}
where $\hat{s}_i(\tau)$ and $v_{\hat{s}_{i}}(\tau)$ denote the estimated mean and variance of $s_i$ in the next step (Process 2) respectively. We let $p_{m\to n}^s(\tau)$ denotes the non-zero possibility of the Bernoulli variable $b_n$ passing from the $m$th sum node to the $n$th variable node in the $\tau$th iteration. The Bernoulli message update of $b_n$ at the $m$th sum node from the $n$th variable node is provided by the message passing from the other variables ${i \in {\cal N}/n}$ to the $m$th sum node. For the $\tau$th iteration, we have
\begin{equation}\label{sum_Ber}
\!p_{m \to n}^s (\!\tau\!)=\!\!\left(\!\!1\!\!+ \!\!\frac{{P\big({b_n} = 0|y_m,u_{m \to n}^s(\tau),v_{m \to n}^s(\tau))}}{{P\big({b_n} = 1|y_m,u_{m \to n}^s(\tau),v_{m \to n}^s(\tau))}} \right)^{-1} \mathop {}\limits_{\mathop {\mathop {}\limits_{} }\limits_{} }\!\!\!.
\end{equation}
In order to prevent the overflow and reduce the computational complexity in above equations, we employ Log-Likelihood Ratios (LLRs) method \cite{LDPC_Guassion,Lei2015_TWC} to replace the update of the non-zero probabilities during the message update process. We have the following LLRs definitions,
\begin{numcases}{}
l_{m\to n}^s(\tau)=\log{\frac{p_{m\to n}^s(\!\tau\!)}{1-p_{m\to n}^s(\!\tau\!)}},\;\; \label{l_1} \\
l_{n\to m}^v(\tau)= \log{\frac{p_{n\to m}^v(\!\tau\!)}{1-p_{n\to m}^v(\!\tau\!)}},\;\; \label{l_2} \\
l_0=\log{\frac{p_0}{1-p_0}}=-\log({{\lambda^{-1}-1}}), \label{l_3} \vspace{0.5mm}
\end{numcases}
for any $n\in {\mathcal{N}}$ and $m\in {\mathcal{M}}$, where $p_0 $ represents the priori probability of $b_n=1$, and $p_0= \lambda$ since we have the priori information $K$.
Plugging (\ref{equ_noi}) and (\ref{sum_Ber}) into  (\ref{l_1}), we have \vspace{1mm}
\begin{equation} \label{2411}
\begin{split}
l_{m \to n}^s(\!\tau\!)  = &-\mathrm{log} \left(\sqrt{\frac{v^{s}_{m\rightarrow n}(\tau)+h^{2}_{mn}v_{\hat{s}_{n}}(\tau)}{v^{s}_{m\rightarrow n}(\tau)}}\right) \\
& -\frac{(y_{m}-u^{s}_{m\rightarrow n}(\tau)-h_{mn}\hat{s}_{n}(\tau))^2}{2(v^{s}_{m \rightarrow n}(\tau)+ h^{2}_{mn }v_{\hat{s}_{n}}(\tau))} \\
& +\frac{(y_{m}-u^{s}_{m \to n}(\tau))^2}{2v^{s}_{m \to n}(\tau)}.
\end{split}
\end{equation}  \vspace{1mm}
\subsubsection{Message Update at Variable Nodes }

Similarly, in right subfigure of the Fig. \ref{fig:All_Factor_Graph_L} (b), we consider the message update in the $n$th variable node  as a broadcast process \cite{Lei2015_TWC,Chongwen_Journal}. Therefore, we have the message update from the  $n$th variable node to the $m$th sum node as,
\begin{equation}\label{24}    
p_{n \to m}^v(\tau\!+\!1)\!= \!\!\frac{{\lambda}\!\!\!\mathop \Pi \limits_{j \in \mathcal{M}/m} \!\!\!p_{j \to n}^s(\!\tau\!)}{{\lambda}\!\!\!\!\mathop \Pi \limits_{j \in \mathcal{M}/m} \!\!\!\!p_{j \to n}^s(\!\tau\!)\! + \!(1 \!-\! \lambda)\!\!\!\!\!\mathop \Pi \limits_{j \in \mathcal{M}/m} \!\!\!\!(\!1\! -\! p_{j \to n}^s(\!\tau\!))},
\end{equation}
where $n\in\mathcal{N}$, and $j \in \mathcal{M}/m$ denotes $m, j \in \mathcal{M}$ and $j\neq m$.
Similarly, we have the LLRs update of the message at the variable node as
\begin{equation}
l_{n \to m}^v(\tau\!+\!1) = l_0 + \!\!\sum \limits_{j\in \mathcal{M}/m}{\!\!{l}_{j\to k}^s}(\!\tau\!),
\end{equation}

Furthermore, we can obtain the estimation of the Bernoulli variable $ b_{n}$ at the $ (\tau+1)$th iteration as
\begin{equation}
\left\{ \begin{array}{l} \vspace{-0mm}
l_{n \to m}^{b}(\tau\!+\!1) = l_0  + \!\!\sum \limits_{j\in \mathcal{M}}{\!\!{l}_{j\to k}^s}(\!\tau\!), \\
\hat{b}_{n}(\!\tau+1\!)=1/(1+e^{-l_{n \to m}^{b}(\!\tau+1\!)}).
\end{array} \right.
\end{equation}

\textbf{\textit{Remark 1:}} $ l^{v}_{m \rightarrow n}(\tau+1) $ is the extrinsic information and will be used to update the mean and variance of the equivalent Gaussian noise $n_{mn}^*(\tau)$ in the next iteration. On the other hand, $l_{n \to m}^{b}$ is the full information updating from all sum nodes. It will be used for updating $\hat{b}_{n}(\tau+1)$  that will be used in the LMMSE Estimation phase for the estimation of $\mathbf{x}$.

\textbf{\textit{Remark 2:}}  The proposed SMP algorithm is\emph{\textbf{ robust to}} the distribution of the sparse signal $\mathbf{x}$ since the sum term of  $\sum_{i\in \mathcal{N}/n} {h_{mi}s_{i}b_i} + n_m$ can be approximated as the Gaussian based on the central limit theorem, regardless of what distribution the sparse signal $\mathbf{x}$ will be.

\subsection{LMMSE Estimation}

 As we all know, the LMMSE estimation is optimal in MSE sense for the linear non-sparse signal. After we obtained the estimation of positions of the non-zero entries in Process 1, the main objective of the Process 2 is to estimate the exact value of the diagonal matrix $\mathbf{D}(\mathbf{s}) $. To reach this goal, one novel estimation method is proposed based on the LMMSE estimation. This method is to \textbf{exchange the locations } between $s_{n}$ and $ b_{n}$ in the (\ref{II3_1}) so that the LMMSE estimator can take full advantage of the estimated sparse information in the Process 1 and obtain an accurate estimation of $\mathbf{x}$. The new LMMSE estimator is given by
\begin{numcases}{}
 \mathbf{\hat{s}}\!=\!\mathbf{V} _{\textbf{\emph{s}}}(\mathbf{HD(\hat{b})})^T \Big((\mathbf{HD(\hat{b})}\mathbf{V} _{\textbf{\emph{s}}}(\mathbf{HD(\hat{b})})^T\!+\!\sigma_{w}^{- 2}\mathbf{I}_M\Big)^{-1} \nonumber  \\
 \quad\quad (\textbf{\emph{y}}-\mathbf{HD(\hat{b})}\mathbf{u}_s) + \mathbf{u}_s,\label{II1}  \\
 \mathbf{\hat{V}}_{\textbf{\emph{s}}} \!= \!(\sigma_{w}^{-2}\mathbf{(HD(\hat{b}))}^T \mathbf{(HD(\hat{b}))}+\mathbf{V}_{\textbf{\emph{s}}}^{-1})^{-1}, \label{II2}
\end{numcases}{}
\!\!where $\mathbf{u}_s$ denotes the mean of $\mathbf{s}$, and $ \mathbf{V}_{\textbf{\emph{s}}} $ and $\mathbf{\hat{V}}_{\textbf{\emph{s}}} \in \mathbb{C}^{N \times N} $ denote the covariance matrix and estimated the covariance matrix of $\mathbf{s}$.  Specifically, the $n$th diagonal element $v_{\hat{s}_n}$ of  $\mathbf{\hat{V}}_{\textbf{\emph{s}}}$ denotes the deviation of the estimation error of the source $s_n$.  

\textbf{\textit{Remark 3:}} The LMMSE estimator ensures that the estimation of non-zero elements of $\mathbf{x}$  is optimal in the MSE sense if there is no sparse information. On the other hand, the AMP-based algorithm cannot make the claim in the same condition since it has the distributive nature of message passing. More important, the proposed novel strategy leverages the estimated sparsity information $\hat{\mathbf{b}}$ that can improve the estimation performance for the sparse signal.
\subsection{ Sparsity Combiner }
When the MSE of the proposed estimation algorithm approaches the minimum or the number of iterations achieves the set requirement, the final estimation of $\mathbf{x}$ was output as

\begin{equation}
\mathbf{\hat{x}}=\mathbf{D}(\mathbf{\hat{s}})\mathbf{\hat{b}}.
\end{equation}

\textbf{\textit{Remark 4:}} It should be pointed out that the final estimation performance is determined by the SMP and LMMSE estimation, and the proposed algorithm takes full use of the both virtues of the SMP and LMMSE estimator.  Moreover, when the channel $\mathbf{x}$ is more sparse, the advantage of the proposed SMP-LMMSE algorithm will becomes more significant. 

\section{ASYMPTOTICAL PERFORMANCE ANALYSIS }
In the following, we present the performance analysis of our proposed SMP-LMMSE algorithm and show that it is asymptotical achievability of the MMSE Bound.
\subsection{MMSE Estimator}
We know that MMSE estimator is theoretically optimal in the MSE sense for the any signal, but it is non-analytical. The MMSE estimator is then defined as the estimator achieving minimal MSE\cite{CRLB_book,MMSE_01} as follows, 
\begin{equation} \label{c1}
\mathbf{u_{x|y}}=\mathrm{arg \,\, min}\; E\{\parallel \mathbf{x}-\mathbf{u_{x|y}}\parallel^2_{2}\},
\end{equation}
and its MSE can be given by the trace of error
\begin{equation} \label{c2}
\begin{split}
\mathbf{MSE_{MMSE}}&=E\{\parallel \mathbf{x}-\mathbf{u_{x|y}}\parallel^2_{2}\} \\
=\int& \int(\mathbf{x}-\mathbf{u_{x|y}})^T(\mathbf{x}-\mathbf{u_{x|y}})f(\mathbf{x},\mathbf{y})d\mathbf{x}d\mathbf{y} \\
=\int& \mathrm{trace}(\mathbf{V}_{\mathbf{x|y}})f(\mathbf{y})d\mathbf{y} \geq 0.
\end{split}
\end{equation}
where $\mathbf{u_{x|y}}$ and $\mathbf{V}_{\mathbf{x|y}}$ denote the MMSE estimation and covariance of $\mathbf{x}$.
\subsection{The Bound of the Proposed Algorithm}

Under the assumption 1, we have the following Lemma 1.

\textbf{\textit{Lemma 1:}}  For the sparse signal, the proposed SMP-LMMSE estimator has better performance than that of LMMSE estimator. This also be denoted by $\mathbf{MSE_{SMP-LMMSE}} \leq \mathbf{MSE_{LMMSE}}$, where $\mathbf{MSE_{LMMSE}}$ and $\mathbf{MSE_{SMP-LMMSE}}$ denote the MSE of LMMSE estimator and SMP-LMMSE estimator.
\begin{IEEEproof}  The MSE of the LMMSE estimator can be obtained by
\begin{equation}\label{49_2}   
\begin{split}
\mathbf{MSE_{LMMSE}} &= E\{\| \mathbf{\hat{x}}-\mathbf{x}\|^{2}_{2}\} \\
=\mathrm{trace}&\{ \mathbf{V_{LMMSE}} \}=\sum\limits_{l= 1}^{N}[\mathbf{V_{LMMSE}}]_{l,l} ,
\end{split}
\end{equation}
where $l \in \{1,2,..,N\}$. Similarly, we can get the MSE of the SMP-LMMSE estimator as follows
\begin{equation}\label{49_3}   
\begin{split}
\mathbf{MSE_{SMP-LMMSE}} &= E\{\| \mathbf{\hat{x}}-\mathbf{x}\|^{2}_{2}\} \\
=\mathrm{trace}\{ \mathbf{V_{SMP-LMMSE}} \}&=\sum\limits_{l= 1}^{N}[\mathbf{V_{SMP-LMMSE}}]_{l,l} ,
\end{split}
\end{equation}
From the assumption 1, we know that the signal vector is $L$ sparse and $\mathbf{V_{SMP-LMMSE}}$ has no more than $L$ eigenvalues. Therefore, the $\mathbf{V_{SMP-LMMSE}}$  can be seen as a singular matrix that obtained from the full rank matrix $\mathbf{V_{LMMSE}}$. The $\mathbf{V_{LMMSE}}$ and $\mathbf{V_{SMP-LMMSE}}$ are the symmetric positive definite matrices since the measurement matrix $\mathbf{H}$ is a non-singular. This means that their all eigenvalues are greater than zero. The eigenvalues of $\mathbf{V_{LMMSE}}$ and $\mathbf{V_{SMP-LMMSE}}$ are denoted as $ 0< \lambda_{N} \leq \lambda_{N-1} \leq ... \leq \lambda_1  $ and  $ 0< \lambda^r_L \leq \lambda^r_{L-1} \leq ... \leq \lambda^r_1 $ respectively. By applying the theorem 4.3.17 in \cite{matrix_Analysis,C_Carbonelli_sparse} obtains
\begin{equation}\label{50}   
\lambda_1 \geq \lambda^r_1 \geq \lambda_2 \geq \lambda^r_2 \geq \cdots  \geq  \lambda_L \geq \lambda^L_L \cdots ,
\end{equation}
and therefore
\begin{equation}\label{51}   
\begin{split}
\mathrm{trace}\{ \mathbf{V_{LMMSE}} \}&= \sum\limits_{l= 1}^{N} \lambda_l \geq \sum\limits_{l= 1}^{L} \lambda^r_l \\
&=\mathrm{trace}\{ \mathbf{V_{SMP-LMMSE}} \}.
\end{split}
\end{equation}
From (\ref{51}), and recalling (\ref{49_2}) and (\ref{49_3}), this shows that  $\mathbf{MSE_{SMP-LMMSE}} \leq \mathbf{MSE_{LMMSE}}$.
\end{IEEEproof}

\textbf{\textit{Lemma 2:}} The MSE of genie-aided MMSE estimator, acts as an lower bound for SMP-LMMSE, i.e.,
\begin{equation}\label{52}   
\mathbf{MSE_{SMP-LMMSE}} \geq \mathbf{MSE_{MMSE}},
\end{equation}
where $\mathbf{MSE_{MMSE}}$ denotes the MSE of genie-aided MMSE estimator.
\begin{IEEEproof}
%
Since the genie-aided MMSE estimator is an idea estimator under the assumption that we have the perfect knowledge of positions of non-zero elements, and the MMSE estimator also minimizes the MSE. Therefore, any implementable estimator will have the MSE greater or equal to the MSE of the MMSE estimator\cite{MMSE,MMSE01,verdo02}. By the definition, we can conclude that the genie-aided MMSE estimator is an lower bound for SMP-LMMSE.
\end{IEEEproof}
\textbf{\textit{Definition 1:}} The Signal Noise Ratio (SNR) of sparse signal $\mathbf{x}$ is defined as
\begin{eqnarray}\label{C3}
\mathrm{SNR}=E\{\parallel \mathbf{x }\parallel_2^{2}\}/E\{\parallel \mathbf{w }\parallel_2^{2}\}.
\end{eqnarray}

\textbf{\textit{Lemma 3:}} When the SNR tends to the infinite or zero, our proposed SMP-LMMSE can achieve the optimal MSE, i.e., the MSE of SMP-LMMSE approaches to that of the genie-aided MMSE bound.
\begin{IEEEproof}
We rewrite (\ref{II1}) and (\ref{II2}) as,
\begin{numcases}{}
\mathbf{\hat{s}}\!=\!(\mathbf{HD(\hat{b})})^T\Big(\mathbf{HD(\hat{b})}(\mathbf{HD(\hat{b})})^T\!+\!\mathrm{SNR}^{-1}\mathbf{I}_M\Big)^{-1} \nonumber \\
\quad\quad (\textbf{\emph{y}}-\mathbf{HD(\hat{b})}\mathbf{u}_s) +\mathbf{u}_s, \label{C4}\\
 \mathbf{\hat{V}} _{\textbf{\emph{s}}} \!= \!\Big(\mathbf{(HD(\hat{b}))}^T\mathbf{V} _{\textbf{\emph{s}}}^{-1}\mathbf{(HD(\hat{b}))}\!+\!\mathrm{SNR}^{-1}\mathbf{V} _{\textbf{\emph{s}}}^{-1}\Big)^{-1} \nonumber \\
 \quad\quad \times \mathrm{SNR}^{-1}. \label{C5}
\end{numcases}{}

When SNR$\rightarrow +\infty$, the above equations can be written as
\begin{numcases}{}
 \mathbf{\hat{s}}=\Big(\mathbf{HD(b)} \Big)^{\dagger}\mathbf{y}, \label{C6} \\
 \mathbf{\hat{V}}_{\textbf{\emph{s}}}= 0. \label{C7}
\end{numcases}{}

The equation (\ref{C6}) and (\ref{C7})  mean that there is no noise and the estimator discards all prior knowledge. Therefore, $\mathbf{MSE_{SMP-LMMSE}}$ became zero at the SNR$\rightarrow +\infty$. Since  $\mathbf{MSE_{MMSE}} \geq 0$ and is the lower bound of $\mathbf{MSE_{SMP-LMMSE}}$, then  $\mathbf{MSE_{MMSE}} =0$ is apparent when SNR$\rightarrow +\infty$. In other words, SMP-LMMSE estimator can be seen as the MMSE estimator when SNR goes to infinite\cite{verdo02}.

On the another case, We rewrite (\ref{II1}) and (\ref{II2}) as, \small {
\begin{numcases}{}
\mathbf{\hat{s}}\!=\!\mathrm{SNR}(\mathbf{HD(\hat{b})})^T \Big(\mathbf{HD(\hat{b})}\mathrm{SNR}(\mathbf{HD(\hat{b})})^T\!+\!\mathbf{I}_M\Big)^{-1} \nonumber\\
\quad\quad (\textbf{\emph{y}}-\mathbf{HD(\hat{b})}\mathbf{u}_s)\!+\!\mathbf{u}_s \label{C8}, \\
 \mathbf{\hat{V}}_{\textbf{\emph{s}}}\!= \!\mathbf{V} _{\textbf{\emph{s}}}-\mathrm{SNR}(\mathbf{HD(\hat{b})})^T \mathbf{HD(\hat{b})}\mathbf{V} _{\textbf{\emph{s}}} \nonumber\\
 \quad\quad \times \Big(\mathbf{HD(\hat{b})}\mathrm{SNR}(\mathbf{HD(\hat{b})})^T\!+\!\mathbf{I}_M\Big)^{-1}. \label{C9}
\end{numcases}{} }

When the SNR$\rightarrow 0$, the above equations become as
\begin{numcases}{}
 \mathbf{\hat{s}}=\mathbf{u}_s, \label{C10} \\
 \mathbf{\hat{V}}_{\textbf{\emph{s}}}= \mathbf{V} _{\textbf{\emph{s}}}. \label{C11}
\end{numcases}{}

The equations (\ref{C10}) and (\ref{C11})  mean that SMP-MMSE estimator ignores the data and only depends on prior information. Apparently, this is also expected at zero SNR for the MMSE estimator. Thus, the SMP-LMMSE setimator also can be seen as MMSE estimator at zero SNR. In summary, when SNR tends to zero or infinite, the MSE of SMP-LMMSE  both tends to  equal that of the MMSE estimator.
\end{IEEEproof}

\textbf{\textit{Assumption 2:}} The elements of the matrix $\mathbf{H}$ are i.i.d. with the distribution $\mathcal{N}(0,1)$. Let $\alpha =L/M$ and $\mathrm{SNR}$ are the fixed constant number.

\textbf{\textit{Proposition 1:}} Assuming that a Genie-aided information provides us with the index set $\mathcal{L}$ of non-zero positions of $\mathbf{x}$,  then,
\begin{equation}\label{C12}
\begin{split}
& \mathbf{MSE_{SMP-LMMSE}}  \\
& \qquad\geq \mathrm{trace}\big({\sigma}_{w}^{-2}(\mathbf{H}_{\mathcal{L}}^T\mathbf{H}_{\mathcal{L}}+SNR^{-1}\mathbf{I}_L)^{-1}\big),
\end{split}
\end{equation}
wherein $\mathbf{H}_{\mathcal{L}} \in \mathbb{C}^{M \times L}$ is the submatrix of $\mathbf{H}$ and its columns corresponding to the index $\mathcal{L}$.
\begin{IEEEproof}
Since we have the Genie-aided information of non-zero positions and $\mathbf{H}_{\mathcal{L}}$ is the submatrix of $\mathbf{H}$, then
$\mathrm{trace}\big(\mathbf{H}_{\mathcal{L}}^T\mathbf{H}_{\mathcal{L}}\big)\leq \mathrm{trace}\big((\mathbf{HD(\hat{b})})^T\mathbf{HD(\hat{b})}\big)$. Combing the assumption 2, the SNR is the fixed, thus we have
\begin{eqnarray}\label{C13}
\begin{split}
 &\mathbf{MSE_{SMP-LMMSE}} \\
&=\mathrm{trace}\big(\sigma_{w}^{-2} (\mathbf{(HD(\hat{b}))}^T \mathbf{(HD(\hat{b}))}+SNR^{-1}\mathbf{I}_M)^{-1}\big)  \nonumber\\
& \geq\mathrm{trace}\big({\sigma}_{w}^{-2}(\mathbf{H}_{\mathcal{L}}^T\mathbf{H}_{\mathcal{L}}+SNR^{-1}\mathbf{I}_L)^{-1}\big).
\end{split}
\end{eqnarray}
\end{IEEEproof}
\textbf{\textit{Lemma 4:}} If the $M \rightarrow \infty$,  then,
\begin{eqnarray}\label{C14}
\begin{split}
&\mathbf{MSE_{SMP-LMMSE}} \\
&\qquad\geq  \mathrm{trace}\big({\sigma}_{w}^{-2}(\mathbf{H}_{\mathcal{L}}^T\mathbf{H}_{\mathcal{L}}+SNR^{-1}\mathbf{I}_L)^{-1}\big)  \\
&\qquad\rightarrow \alpha \sigma_{w}^{-2}.
\end{split}
\end{eqnarray}
\begin{IEEEproof}
Define $\mathbf{A} \triangleq \mathbf{H}_{\mathcal{L}}^T\mathbf{H}_{\mathcal{L}} $. Then, we have
\begin{eqnarray}\label{C15}
[\mathbf{A}]_{k,l} = \sum_{j=1}^{M} [\mathbf{H}_{\mathcal{L}}]_{j,k}^T[\mathbf{H}_{\mathcal{L}}]_{j,l}
\end{eqnarray}
According to the assumption 2, we know that the entries of $\mathbf{H}$ are modeled as i.i.d. with the distribution $\mathcal{N}(0,1)$, we take full advantage of the law of large number as follows. When $M \rightarrow \infty$ and $k \neq l$ \cite{CRB_compressed_sensing,verdo01}, we have
\begin{eqnarray}\label{C16}
\begin{split}
\frac{1}{M}[\mathbf{A}]_{k,l} & =\frac{1}{M}\sum_{j=1}^{M} [\mathbf{H}_{\mathcal{L}}]_{j,k}^T[\mathbf{H}_{\mathcal{L}}]_{j,l} \\
& \rightarrow E\{[\mathbf{H}_{\mathcal{L}}]_{j,k}^T\}E\{[\mathbf{H}_{\mathcal{L}}]_{j,l}\}=0,
\end{split}
\end{eqnarray}
and
\begin{eqnarray}\label{C17}
\begin{split}
\frac{1}{M}[\mathbf{A}]_{k,k}&=\frac{1}{M}\sum_{j=1}^{M} [\mathbf{H}_{\mathcal{L}}]_{j,k}^T[\mathbf{H}_{\mathcal{L}}]_{j,k} \\
&\rightarrow E\{|[\mathbf{H}_{\mathcal{L}}]_{j,k}|^2\}=1,
\end{split}
\end{eqnarray}
Therefore,  we have
\begin{eqnarray}\label{C18}
\frac{1}{M}\mathbf{A}=\frac{1}{M} \mathbf{H}_{\mathcal{L}}^T\mathbf{H}_{\mathcal{L}} \xrightarrow {M \rightarrow \infty} \mathbf{I}_L.
\end{eqnarray}
We note that $\mathbf{A}/M$ is non-singular with the probability 1, and it can be denoted by the $\frac{1}{M} \mathbf{H}_{\mathcal{L}}^T\mathbf{H}_{\mathcal{L}}$. From the result in (\ref{C17}), we have the limit $\lim\limits_{M \to \infty}(\mathbf{A} /M)^{-1}$ exists with the probability 1. Hence, when $M \rightarrow \infty$, we have
\begin{eqnarray}\label{C19}
\begin{split}
&\mathbf{MSE_{SMP-LMMSE}} \\
& \qquad\geq \mathrm{trace}\Big(\frac{{\sigma}_{w}^{-2}}{M}\big(\frac{\mathbf{H}_{\mathcal{L}}^T\mathbf{H}_{\mathcal{L}}}{M}+\frac{\mathrm{SNR}^{-1}\mathbf{I}_L}{M}\big)^{-1}\Big).
\end{split}
\end{eqnarray}
Since $\alpha =L/M$ and $\mathrm{SNR}$ are both the fixed constant number, and plug (\ref{C18}) into the (\ref{C19}),  we have
\begin{eqnarray}\label{C20}
\begin{split}
\mathrm{trace}\Big(\frac{{\sigma}_{w}^{-2}}{M}\big(\frac{\mathbf{H}_{\mathcal{L}}^T\mathbf{H}_{\mathcal{L}}}{M}+\frac{\mathrm{SNR}^{-1}\mathbf{I}_L}{M}\big)^{-1}\Big) \xrightarrow {M \rightarrow \infty} \alpha {\sigma}_{w}^{-2}.
\end{split}
\end{eqnarray} \vspace{-0mm}
Thus, the (\ref{C14}) is proved.
\end{IEEEproof}
\vspace{-1mm}
\section{NUMERICAL RESULTS}

\begin{figure}
\begin{center}
  \includegraphics[width=84mm]{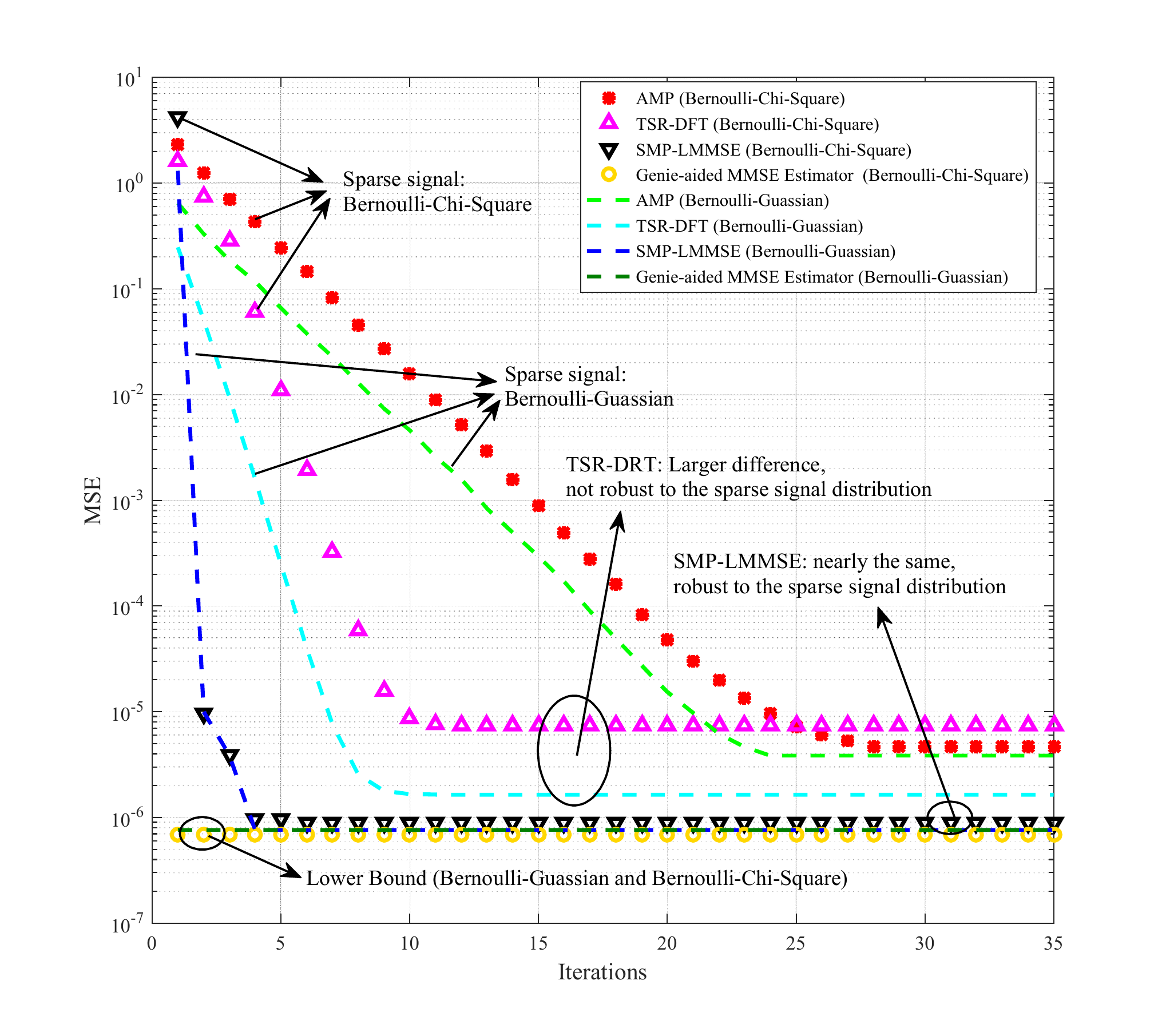} \vspace{-4mm}
  \caption{Comparisons of the proposed SMP-LMMSE, TSR-DFT, AMP and the Genie-aided MMSE estimator under different sparse signal distributions, i.e., Bernoulli-Gaussian and Bernoulli-Chi-Square distribution.  $SNR=50dB$, $N=8192$, $M=4096 \;(=0.5N)$, and $ \lambda =0.125 $. }\label{f2}
\end{center} \vspace{-6mm}
\end{figure}\vspace{-0mm}
We now present the results of a numerical simulation on sparse signal estimation, where the main objective was to estimate the sparse signal ($\mathbf{x} \in \mathbb{C}^{N \times 1} $) from the measurements $\mathbf{y}$ according to the equation (\ref{I1}). Note that the proposed estimation algorithm can be used in many wireless communication scenarios directly. In this section, we demonstrate using the channel estimation for a mmWave MIMO communication system, in which the sparse signal $\mathbf{x}$ denotes the frequency-domain impulse responses of the mmWave MIMO channel, and we consider that the system is a narrow band and channel responses are frequency-flat, and $N$, $M$, $\mathbf{H}$, and $\mathbf{y}$ denotes the number of transmit antennas, the length of the training sequences, known training matrix, and the observed signal vector respectively. In detail, we use the $N=8192$, $M=4096\;(=0.5N)$, and $\mathbf{x}$ under two cases of the Bernoulli-Gaussian and Bernoulli-Chi-Square distribution. The non-zero entries of these two distributions were drew with  $\mathcal{N}(0,1)$ and 4 degrees of freedom respectively, and their indices were generated by the independent and random way. All simulations were run 10000 times, and the averaged results are reported.

In Fig. 3, we show the MSE performance of the proposed estimator versus several typical estimators for the sparse signal $\mathbf{x}$ that follows different distributions, i.e., Bernoulli-Gaussian and Bernoulli-Chi-Square distribution. These typical estimators involves TSR-DFT, Genie-aided MMSE estimator (idea estimator), and AMP with i.i.d. Gaussian measurement matrices. The implementations of AMP and TSR-DFT estimator were based on \cite{rangan} and \cite{majunjie_01} respectively.  We set the $SNR=50dB$ and $ \lambda =0.125 $ in simulations.  It can be seen that the proposed SMP-LMMSE estimator converges faster than AMP and TSR-DFT under two sparse signal types. This is because the convergence speed of the proposed algorithm is only determined by the first process. From the simulations, we observe that the first process usually only needs 4-6 iterations to converge, therefore, the proposed algorithm converges faster. Then, we also can be seen that SMP-LMMSE also reaches the lower MSE and finally converges to the lower bound, i.e., the MSE of the MMSE estimator. Moreover, the Bernoulli-Chi-Square signal deteriorates the MSE performance of TSR-DFT about $7 dB$ in comparison with the Bernoulli-Gaussian. Similarly, estimating the Bernoulli-Chi-Square needs four more iterations to obtain the stable convergence for AMP. On the other hand, the MSE and convergence rate of the proposed SMP-LMMSE are nearly the same for the estimation of Bernoulli-Gaussian and Bernoulli-Chi-Square signal. These means that the proposed SMP-LMMSE algorithm is \textbf{robust }to the distribution of non-zero entries of the sparse signal $\mathbf{x}$.

Fig. 4, shows the numerical results for the analytical lower bound, upper bound versus our proposed algorithm in the different iterations. First, we see that the simulation and analytical results (Lemma 1 and Lemma 2) for the proposed algorithm agree very well. Note that only simulation results are provided for Genie-aided MMSE estimator since there is no implementable technique for archiving the optimal MMSE. From Fig. 4, we see that the MSE of SMP-LMMSE gradually converges to the lower bound (MMSE bound) when the SNR tends to the infinity or zero, which also verifies the Lemma 3.

\begin{figure}
  \centering
  \includegraphics[width=88mm]{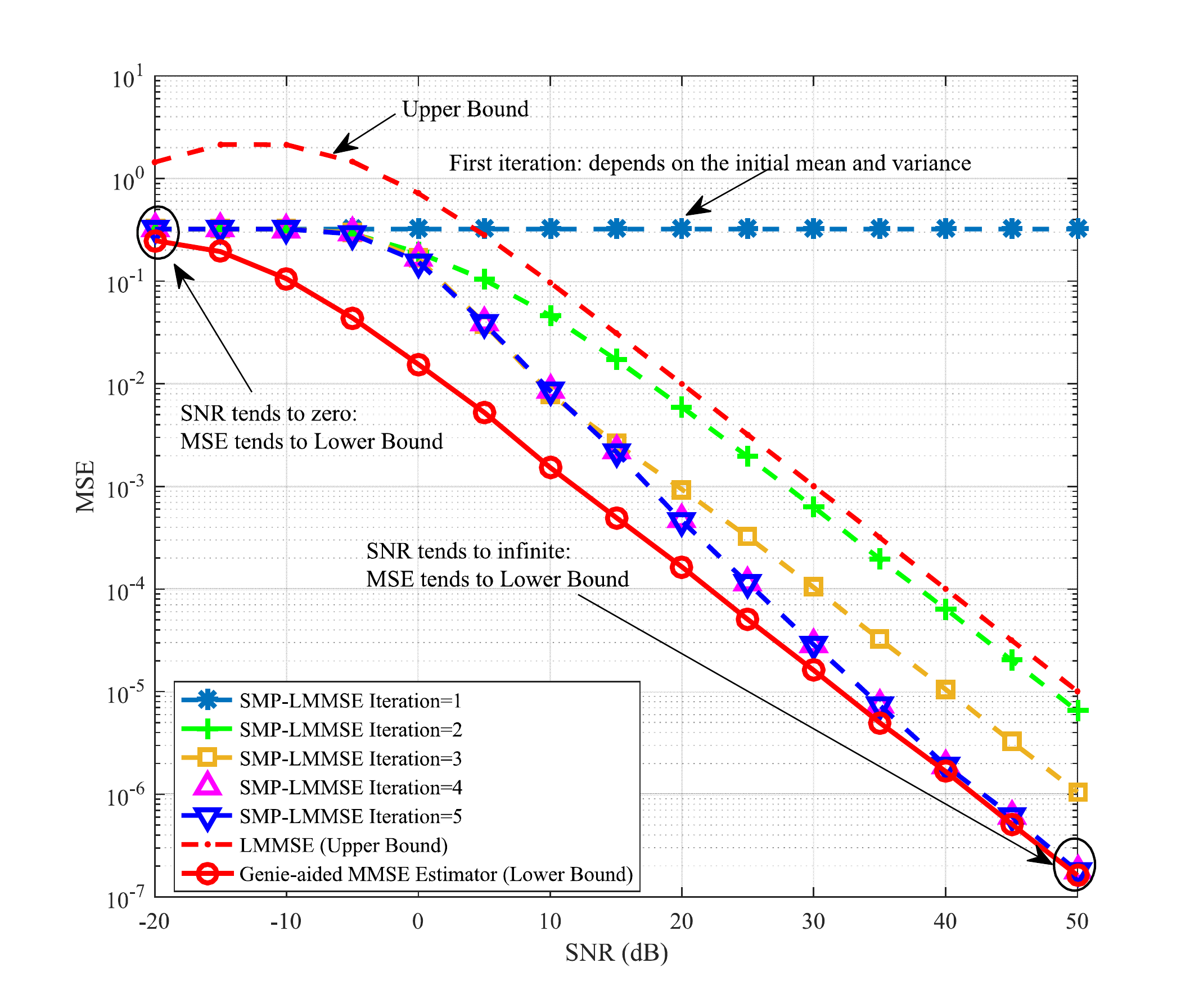}\vspace{-4mm}\\
  \caption{ Comparison of the analytical lower bound, upper bound vs the our proposed algorithm in the different iterations. $N=8192$, $M=4096$, $ \lambda =0.04 $, and $SNR=\{-20,-10,0,10,20,30,40,50\}dB$. }\label{f2} \vspace{-6mm}
\end{figure}

\section{CONCLUSION}

An asymptotically optimal estimator for the sparse signal was proposed, which is robust to the distribution of non-zero elements of the sparse signal $\mathbf{x}$, and it can be used in many wireless communication scenarios directly. Moreover, the performance upper bound and lower bound of the proposed estimator are also analysed. In the low or high SNR regime, simulation results agree with the analysis very well and shows that the proposed algorithm asymptotically converges the optimal performance (Genie-aided MMSE estimator). More important, these numerical results illustrate that the SMP-LMMSE algorithm outperforms the AMP and TSR-DFT algorithm, and is also more robust to the distribution of the sparse signal. It is an interesting future research topic to develop the state evolution analysis of the proposed algorithm and establish the justifications for the state evolution.

\end{document}